\documentstyle[twocolumn]{jpsj}
\def\runtitle{Ferrimagnetic Spin Chains}
\def\runauthor
{Shoji {\sc Yamamoto} and T$\hat{\mbox o}$ru {\sc Sakai}}

\title
{Low-Energy Structure of Heisenberg Ferrimagnetic Spin Chains}

\author
{Shoji {\sc Yamamoto} and T$\hat{\mbox o}$ru {\sc Sakai}$^*$}

\inst
{Department of Physics, Faculty of Science, Okayama University,\\
 Tsushima, Okayama 700-8530\\
 $^*$Faculty of Science, Himeji Institute of Technology,\\
 Kamigori, Ako, Hyogo 678-1297}

\recdate{21 August 1998}

\abst
{Static and dynamic structure factors of Heisenberg ferrimagnetic
 spin chains are numerically investigated.
 There exist two distinct branches of elementary excitations, which
 exhibit ferromagnetic and antiferromagnetic aspects.
 The ferromagnetic feature is smeared out with the increase of
 temperature, whereas the antiferromagnetic one persists up to
 higher temperatures.
 The scattering intensity is remarkably large at lower boundaries
 of the ferromagnetic and antiferromagnetic spectra.
 All these observations are consistent with the
 ferromagnetic-to-antiferromagnetic crossover in the thermal
 behavior which has recently been reported.}

\kword
{ferrimagnetic spin chain,
 static structure factor, dynamic structure factor,
 quantum Monte Carlo, exact diagonalization}

\begin{document}
\sloppy
\maketitle

   Recently considerable attention has been directed to quantum
magnets with two kinds of antiferromagnetically exchange-coupled
centers.
Since coexistent spins of different kinds do not necessarily result
in magnetic ground states, it remains a stimulating problem whether
the system is massive or massless.
For example, an alignment of elementary cells with spins $1$, $1$,
$1/2$, and $1/2$ in this order results in a singlet ground state and
an excitation gap immediately above it.\cite{Tone00}
Several authors \cite{Fuku09,Fuku98,Koga22,Taka55} have generally
discussed what is the criterion for the massive phases via the
nonlinear-$\sigma$-model technique.
On the other hand, the most simple alternating-spin chains
\cite
{Alca67,Pati94,Kole36,Breh21,Nigg31,Yama09,Ono76,Kura29,Yama08,Ivan14,Mais}
with two kinds of spins $S$ and $s$ which are described by the
Hamiltonian
\begin{equation}
   {\cal H}
      =J\sum_j
        \left(
         \mbox{\boldmath$S$}_{j} \cdot \mbox{\boldmath$s$}_{j}
        +\mbox{\boldmath$s$}_{j} \cdot \mbox{\boldmath$S$}_{j+1}
        \right)\,,
   \label{E:H}
\end{equation}
show ferrimagnetism instead of antiferromagnetism.
Considering that all the mixed-spin-chain compounds synthesized so
far \cite{Kahn95} exhibit ferrimagnetic ground states, the study on
the model (\ref{E:H}), the simplest example of a quantum ferrimagnet,
is all the more important.
Although pioneering authors \cite{Dril13} have already discussed
theoretically quantum ferrimagnets in the $1980$s, their interest was,
for instance, the quantitative dependence of the thermal behavior on
constituent spin quantum numbers.
The exact diagonalization of short chains was not necessarily
conclusive in understanding the universal quantum ferrimagnetic
behavior especially at low temperatures.
Recent arguments
\cite
{Alca67,Pati94,Kole36,Breh21,Nigg31,Yama09,Ono76,Kura29,Yama08,Ivan14,Mais}
have renewed the interest in this fascinating subject and have even
motivated brand-new NMR measurements \cite{Fuji} which are currently
in progress.

   Let us consider the Hamiltonian (\ref{E:H}) under the periodic
boundary condition.
We set the number of unit cells to $N$ and assume that $S>s$.
Then the Lieb-Mattis theorem \cite{Lieb49} immediately shows that
the Hamiltonian (\ref{E:H}) has ferrimagnetic ground states of spin
$(S-s)N$.
As we may generally expect gapless excitations from magnetic
ground states, we here take little interest in the naive problem of
whether the spectrum is gapped or gapless.
Actually, gapless excitation branches \cite{Yama09} lie in the
subspaces whose total spin is less than $(S-s)N$, showing
quadratic dispersion relations at small momenta.
Thus the low-lying excitations of quantum ferrimagnets have a
ferromagnetic aspect.
The other excitation branches \cite{Yama09} lying in the subspaces
whose total spin is greater than $(S-s)N$ are all gapped and
reminiscent of gapped antiferromagnets.\cite{Hald64}

   The thus-revealed low-energy structure may be recognized as a
combination of ferromagnetic and antiferromagnetic features
\cite{Yama08} and is well explained in terms of the spin-wave
theory.\cite{Pati94,Breh21}
For the sake of argument, we briefly review the spin-wave treatment.
We introduce the bosonic operators for the spin deviation from a
N\'eel state with $M\equiv\sum_j(S_j^z+s_j^z)=(S-s)N$ as
\begin{equation}
   \left.
   \begin{array}{ccc}
      S_j^+=\sqrt{2S}\,a_j\,,&
      S_j^z=S-a_j^\dagger a_j\,,\\
      s_j^+=\sqrt{2s}\,b_j^\dagger\,,&
      s_j^z=-s+b_j^\dagger b_j\,.
   \end{array}
   \right.
   \label{E:HPBoson}
\end{equation}
Setting the lattice constant to $a$, we define the momentum
representation of the bosonic operators as
\begin{equation}
   \left.
   \begin{array}{c}
      a_k={\displaystyle \frac{1}{\sqrt{N}}}
          {\displaystyle \sum_j}
          {\rm e}^{ 2{\rm i}ak(j-1/4)}a_j\,,\\
      b_k={\displaystyle \frac{1}{\sqrt{N}}}
          {\displaystyle \sum_j}
          {\rm e}^{-2{\rm i}ak(j+1/4)}b_j\,.\\
   \end{array}
   \right.
   \label{E:FourierB}
\end{equation}
The Bogoliubov transformation
\begin{equation}
   \left.
   \begin{array}{c}
      \alpha_k=\mbox{cosh}\theta_k\ a_k
              +\mbox{sinh}\theta_k\ b_k^\dagger\,,\\
      \beta_k =\mbox{sinh}\theta_k\ a_k^\dagger
              +\mbox{cosh}\theta_k\ b_k\,,
   \end{array}
   \right.
   \label{E:Bogoliubov}
\end{equation}
with
\begin{equation}
   \mbox{tanh}2\theta_k
     =-\frac{2\sqrt{Ss}}{S+s}
      \mbox{cos}(ak)\,,
   \label{E:theta}
\end{equation}
results in the two distinct dispersion relations
\begin{equation}
   \omega_{k}^\mp=\omega_k\mp(S-s)\,,
   \label{E:omegamp}
\end{equation}
where
\begin{equation}
   \omega_k=\sqrt{(S-s)^2+4Ss\sin^2(ak)}\,.
   \label{E:omega}
\end{equation}
The $\alpha$- and $\beta$-bosons, respectively, have the effects of
reducing and enhancing the ground-state magnetization and may
therefore be regarded as ferromagnetic and antiferromagnetic.
The obtained dispersions $\omega_k^-$ and $\omega_k^+$ are indeed
reminiscent of those of ferromagnets and gapped antiferromagnets,
respectively, though the antiferromagnetic gap within the
lowest-order spin-wave theory, $\omega_{k=0}^+=J$, is much smaller
than the true value ${\mit\Delta}=1.759J$.\cite{Breh21,Yama09}

   Two distinct features are most prominently exhibited in the
temperature dependence of the specific heat.\cite{Yama08}
At low temperatures a ferromagnetic $T^{1/2}$ initial behavior
appears, whereas at intermediate temperatures we encounter a
Schottky-like peak which can simply be described by the
antiferromagnetic gap ${\mit\Delta}$.
Motivated by these observations, we here investigate the low-energy
structure of quantum ferrimagnets in detail.
Although the low-lying antiferromagnetic excitations lie in the
background of the ferromagnetic spectra, why does such a
well-pronounced Schottky-like peak appear?
This is the subject in the present article.
Since previous numerical investigations \cite{Alca67,Pati94} suggest
that quantum behavior of the model (\ref{E:H}) is qualitatively the
same regardless of the values of $S$ and $s$ as long as they differ
from each other, we restrict our argument to the case of
$(S,s)=(1,1/2)$.

   In order to reveal how the ferromagnetic and antiferromagnetic
aspects are exhibited, let us investigate static structure factors
as functions of temperature.
Generally, the static structure factors of ferromagnetic mono-spin
chains have their peaks at the zone center $k=0$, while those of
antiferromagnetic mono-spin chains at the zone boundary $k=\pi/a$.
However, in the present case, the system is composed of two
interacting sublattices and both ferromagnetic and
antiferromagnetic peaks appear at the center of the reduced
Brillouin zone $-\pi/2a<k\leq\pi/2a$.
Therefore the position of the peak can not be used to determin its
nature.
In an attempt to find relevant operators to detect the ferromagnetic
and antiferromagnetic modes, we here rely on the spin-wave theory,
namely, we regard the $\alpha$-boson and $\beta$-boson correlations
as ferromagnetic and antiferromagnetic, respectively.
Keeping in mind that the present model is isotropic, thermal
averages
$\langle\alpha_k^\dagger\alpha_k\rangle$ and
$\langle\beta_k^\dagger\beta_k\rangle$
can be calculated in terms of spin operators as
\begin{equation}
   \left.
   \begin{array}{c}
   \langle\alpha_k^\dagger\alpha_k\rangle
    \equiv S^-(k)
    =\langle O_{-k}^-O_{k}^-\rangle\,,\\ 
   \langle\beta_k^\dagger\beta_k\rangle
    \equiv S^+(k)
    =\langle O_{-k}^+O_{k}^+\rangle\,,\\
   \end{array}
   \right.
\end{equation}
with
\begin{equation}
   \left.
   \begin{array}{c}
      O_k^-=          \mbox{cosh}\theta_k\ S_k^z
           +\sqrt{2}\,\mbox{sinh}\theta_k\ s_k^z\,,\\
      O_k^+=          \mbox{sinh}\theta_k\ S_k^z
           +\sqrt{2}\,\mbox{cosh}\theta_k\ s_k^z\,,\\
   \end{array}
   \right.
   \label{E:Okpm}
\end{equation}
where the Fourier transforms of the spin operators are generally
defined as
\begin{equation}
   \left.
   \begin{array}{c}
      S_k^\lambda
        ={\displaystyle \frac{1}{\sqrt{N}}}
         {\displaystyle \sum_j}
         {\rm e}^{2{\rm i}ak(j-1/4)}S_j^\lambda\,,\\
      s_k^\lambda
        ={\displaystyle \frac{1}{\sqrt{N}}}
         {\displaystyle \sum_j}
         {\rm e}^{2{\rm i}ak(j+1/4)}s_j^\lambda\,,\\
   \end{array}
   \right.
   \label{E:FourierS}
\end{equation}
and $\theta_k$ is defined in eq. (\ref{E:theta}).

   Employing a quantum Monte Carlo method,\cite{Breh21,Yama64}
we have calculated the thus-defined static structure factors at
various temperatures, which are shown in Fig.~\ref{F:static}.
Both structure factors $S^{\mp}(k)$ have their peaks at $k=0$, which
reflects a combination of ferromagnetic and antiferromagnetic
features in this system.
However, they show quantitatively different temperature dependences.
While the ferromagnetic peak is rapidly smeared out as the
temperature increases, the antiferromagnetic one still persists at
rather high temperatures.
This observation implies that the antiferromagnetic correlation
is dominant at intermediate temperatures and thus well explains
the well-pronounced Schottky-like peak of the specific
heat.\cite{Yama08}

   The existence of a gapped antiferromagnetic mode does not
necessarily result in a Schottky-like behavior of the specific
heat, which is basically inherent in two-level systems.
As long as the specific heat shows a prominent peak reflecting the
gap ${\mit\Delta}$, the lowest-lying antiferromagnetic mode should be
fully stressed in the spectrum.
In this context we inquire further into dynamic structure factors.
We introduce the ferromagnetic and antiferromagnetic dynamic
structure factors at $T=0$ as
\begin{equation}
   \left.
   \begin{array}{c}
   S^{-+}(k,\omega)
     =\sum_n
      \vert\langle n\vert S_k^- +s_k^- \vert 0\rangle\vert^2
      \delta\big(\omega-(E_n-E_0)\big)\,,\\
   S^{+-}(k,\omega)
     =\sum_n
      \vert\langle n\vert S_k^+ +s_k^+\vert 0\rangle\vert^2
      \delta\big(\omega-(E_n-E_0)\big)\,,\\
   \end{array}
   \right.
\end{equation}
respectively, where
$S_k^\pm=S_k^x\pm{\rm i}S_k^y$,
$s_k^\pm=s_k^x\pm{\rm i}s_k^y$,
$\vert n\rangle$ denotes an eigenstate of the Hamltonian (\ref{E:H})
with energy $E_n$, and $E_0$ is set to the ground-state energy.
The ground states of quantum ferrimagnets are macroscopically
degenerate and thus the choice for $\vert 0\rangle$ is not unique.
We here assume the highest-magnetization ground state, namely, the
lowest-energy state with $M=(S-s)N$, as the initial state
$\vert 0\rangle$ so that $S^{-+}(k,\omega)$ and $S^{+-}(k,\omega)$
can detect the ferromagnetic and antiferromagnetic elementary
excitations, respectively.
$S^{\sigma\bar\sigma}(k,\omega)$
($\sigma=\pm$, $\bar\sigma=-\sigma$)
is generally expressed in terms of its corresponding Green's
function as
\begin{equation}
   S^{\sigma\bar\sigma}(k,\omega)
     =-\frac{1}{\pi}{\rm Im}\,G^{\sigma\bar\sigma}(k,\omega)\,.
\end{equation}
$G^{\sigma\bar\sigma}(k,\omega)$ is given as a continued fraction
\cite{Gagl99}
\begin{equation}
   G^{\sigma\bar\sigma}(k,\omega)
     =\frac
      {\langle 0\vert
       S_{-k}^{\bar\sigma}S_{k}^{\sigma}
       \vert 0\rangle}
      {\omega-a_0-\frac
                  {b_1^2}
                  {\omega-a_1-\frac
                              {b_2^2}
                              {\omega-\cdots}}}\,,
\end{equation}
where the coefficients $a_n$ and $b_n$ are obtained by the Lanczos
method through a recursive relation
\begin{equation}
   \left.
   \begin{array}{c}
   \vert f_{n+1}\rangle
     ={\cal H}\vert f_{n}\rangle
     -a_n\vert f_{n}\rangle
     -b_n^2\vert f_{n-1}\rangle\,,\\
   \vert f_{0}\rangle=S_k^\sigma\vert 0\rangle\,,\ \
   b_0=0\,,\\
   \end{array}
   \right.
\end{equation}
with a set of orthogonal states $\{\vert f_{n}\rangle\}$.
This method works fairly well \cite{Haas81,Taka45} for the study of
low-lying states of various one-dimensional magnets.

   The thus-calculated dynamic structure factors are shown in Fig.
\ref{F:dynamic}.
We find qualitatively the same behavior at all the chain lengths we
treated and therefore the present observations may be regarded as
the long-chain behavior.
This is due to the considerably small correlation length
\cite{Pati94,Breh21} of the system.
In effect, the antiferromagnetic gap ${\mit\Delta}$ is $1.7594J$,
$1.7592J$, and $1.7591J$ at $N=8$, $N=10$, and $N=12$, respectively,
the values of which are almost seen to converge.
The scattering intensity is remarkably large at the lower boundaries
of the ferromagnetic and antiferromagnetic spectra at each $k$, as
was expected.
In particular the lowest-lying antiferromagnetic poles hold most of
the antiferromagnetic scattering weight.
They are never smeared out in the ferromagnetic spectra but are
prominently present.
As far as antiferromagnetic eigenvalues are concerned, we can in
principle detect all of them in the subspace of $M=(S-s)N+1$, and
therefore, their lower boundary seems to form a split band
separated from an upper continuum.
Thus the low-energy structure is essentially described by these
well-pronounced ferromagnetic and antiferromagnetic bands, which
are almost parallel to each other and are separated by the gap
${\mit\Delta}=1.759J$.
This explains the reason why the specific heat is well fitted to the
simple Schottky form \cite{Yama08}
\begin{equation}
   \frac{C}{Nk_{\rm B}}
     \propto
      \left(
       \frac{\mit\Delta}{2k_{\rm B}T}
      \right)^2
      {\rm sech}^2
      \left(
       \frac{\mit\Delta}{2k_{\rm B}T}
      \right)\,,
\end{equation}
at intermediate and even higher temperatures.

   In conclusion, we have numerically studied the low-energy
structure of Heisenberg ferrimagnetic spin chains in an attempt to
understand their novel thermal behavior.\cite{Yama08}
The two distinct static structure factors illustrate why the clear
ferromagnetic-to-antiferromagnetic crossover is observed in the
temperature dependence of the thermodynamic quantities.
As the temperature increases, the ferromagnetic nature rapidly fades
out, whereas the antiferromagnetic one persists up to higher
temperatures.
That is why the specific heat exhibits a well-pronounced
Schottky-like peak and the susceptibility-temperature product shows
an increasing behavior with temperature, in spite of the
low-temperature ferromagnetic behavior.
On the other hand, the dynamic structure factors clearly indicate
that the present model effectively behaves as a two-level system
unless the temperature is sufficiently low.
The findings are a suggestive guide for neutron-scattering
measurements.
Recently Hagiwara {\it et al.} \cite{Hagi09} have succeeded in
obtaining the fundamental parameters of a bimetallic chain
compound NiCu(pba)(D$_2$O)$_3$$\cdot$$2$D$_2$O, where
pba = 1,3-propylenebis(oxamato), which is a prototype of the
$(S,s)=(1,1/2)$ ferrimagnet.
We hope that our calculations will motivate
further experimental study on this topic.

   The authors would like to thank N. Fujiwara and M. Hagiwara for
their useful comments.
This work was supported by the Japanese Ministry of Education,
Science, and Culture through a Grant-in-Aid (No. 09740286)
and by the Okayama Foundation for Science and Technology.
Part of the numerical computation was carried out using the facility
of the Supercomputer Center, Institute for Solid State Physics,
University of Tokyo.

\begin{figure}
\caption{Ferromagnetic (a) and antiferromagnetic (b) static structure
         factors as functions of temperature for the chain of
         $N=32$.
         The temperature is indicated in units of $J/k_{\rm B}$ in
         the inset.}
\label{F:static}
\end{figure}

\begin{figure}
\caption{Ferromagnetic ($\times$) and antiferromagnetic ($+$)
         dynamic structure factors at the absolute zero temperature.
         The lowest-energy five poles are shown for each of them.
         The intensity of each pole is proportional to the area of the
         circle.}
\label{F:dynamic}
\end{figure}

\end{document}